# Switchable two-state plasmonic tweezers for dynamic manipulation of nano-objects

*Gabriele C. Messina[1,2*], Xavier Zambrana-Puyalto[1], Nicolò Maccaferri[1], Denis Garoli[1], Francesco De Angelis[1*]*

[1]*Plasmon Nanostructures, Istituto Italiano di Tecnologia, Via Morego 30, 16163 Genova GE, Italy*

[2]*Present Address: Intense Laser Irradiation Laboratory (ILIL), Istituto Nazionale di Ottica (CNR-INO), Via Giuseppe Moruzzi 1, 56124 Pisa PI, Italy*

*Corresponding author: gabriele.messina@ino.it; francesco.deangelis@iit.it*

**Abstract**

In this work we present a plasmonic platform capable of trapping nano-objects as small as 100 nm in two different spatial configurations. The switch between the two trapping states, localized on the tip and on the outer wall of a vertical gold nanochannel, can be activated by a variation in the focusing position of the excitation laser along the main axis of the nanotube.

We show that the trapping mechanism is facilitated by both an electromagnetic and thermal action.

The "inner" and "outer" trapping states are respectively characterized by a static and a dynamic behavior and their stiffness was measured by analyzing the position of the trapped specimens as a function of time. In addition, it was demonstrated that the stiffness of the static state is high enough to trap of particles as small as 40nm.

These results show a simple, controllable way to generate a switchable two-state trapping regime, which could find applications as a model for the study of dynamic trapping or as mechanism for the development of nanofluidic devices.

**Introduction**

Since its first demonstration in the 1980s [1], the versatility of the optical tweezing phenomenon has drawn the attention of different scientific communities. Among others, optical tweezers find application both as a model for studies on optomechanics [2] and as a tool for nano- and biotechnology [3,4,5].

However, common tweezing approaches suffer from disadvantages related to the need of bulky and expensive high numerical aperture objectives. Furthermore, classic optical trapping displays substantial difficulties when dealing with objects with dimensions of tens of nanometers.

In this regard, the combination of conventional tweezing methods with plasmonics, which offers the possibility to concentrate the electromagnetic (EM) field in small volumes (the so-called hot-spots), allowed to expand the range of their application to the nanoscale. Moreover, the intense EM field enhancement induced by metal nanostructures promotes an increase in the depth of the trapping potential. This increased stability reduces the requirement for high laser powers, thus avoiding the possibility of temperature rise, which potentially leads to convection as well as specimen damage.

Usually, plasmonic tweezing experiments can be performed in flow-over or flow-through configurations.

Flow-over approaches are based on the use of dielectric substrates decorated with planar nanoantennas. The antennas are illuminated with a microscope objective [6] or by a collimated laser beam to respectively induce the trapping of single or multiple nanoobjects [7,8,9].

These arrangements allow to tweeze a large number of particles at the same time, but suffer from limitations commonly related to flow-over layouts.

Indeed, when working with diluted solutions, the statistical possibility to trap molecules in the hot-spot becomes strongly correlated to the diffusion probability inside the solution [10].

To overcome this limitation, recent works have introduced the use of thermal induced convective forces [11,12,13]. Nevertheless, trapped particles could stick on the surface of the substrate, thus compromising the reproducibility of experiments.

An alternative is represented by the use of flow-through systems, such as plasmonic nanopores drilled on gold-covered $Si_3N_4$ membranes [14].

These structures show very good trapping performances, which can be further improved by including plasmonic hot-spots in the nanostructure design. The resulting geometries, such as double nanoholes [15] or inverted bowties [16], are characterized by gaps of a few tens of nanometers in which the electromagnetic field is strongly confined and enhanced, leading to efficient trapping states.

However, the performance of flow-through arrangements is constrained by the transverse section of the structure, as the flow is proportional to it.

Beside the problems related to the fluidic configuration, the integration of plasmonic tweezing into lab-on-chip devices has also been slowed down by the limited manipulation capabilities of the technique.

Literature reports some examples of dynamic manipulation of particle based on plasmonic nanostructures [17,18]. These approaches are mostly characterized by binary on/off configurations, where the particle can be either trapped by the nanostructure-enhanced field (on) or released in the solution (off). In these conditions, the possibility to trap the same particle multiple times is therefore a stochastic event and, when working with very diluted media, it could represent a time consuming option.

Here we present a novel approach for the stable trapping of nanosized objects in two spatially separated arrangements whose selective activation can be triggered by a change in the excitation configuration of the structure.

Our system is based on the use of tridimensional plasmonic nanochannels stemming out of a planar substrate.

The field distribution along the nanoantennas shows two distinct maxima, respectively located at the tip and on the outer wall of the channel, with field intensities high enough to allow for the trapping of nanometric objects on both zones.

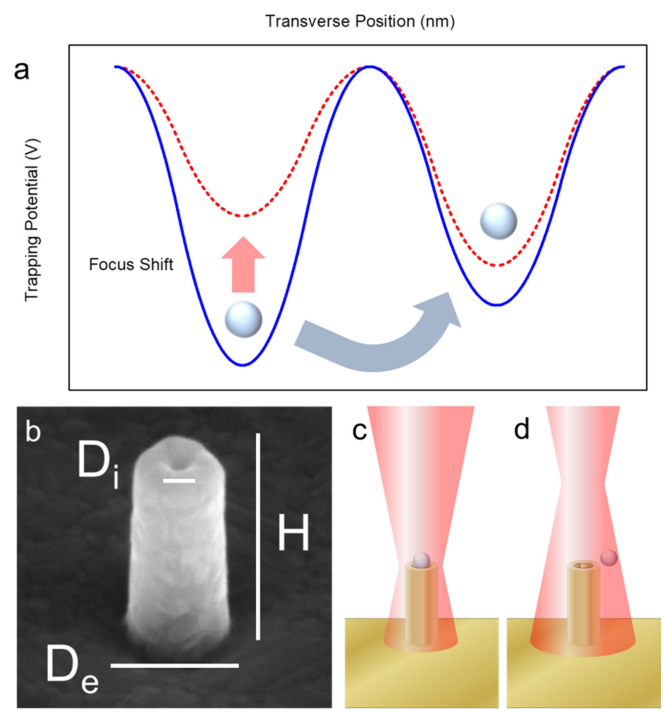

Figure 1. (a) Scheme of the trapping potential distribution as a function of the excitation laser focus position on a gold vertical nanoantenna. (b) Scanning Electron Microscopy image of a nanotube used for the trapping experiments (H=400 nm, $D_e$= 180 nm, $D_i$= 40 nm). A shift in the focus position from the tip of the antenna to 2 μm above the nanotube results in the trapping of beads respectively in correspondence of the inner channel (c) and of the outer ring (d).

As sketched in Fig.1a, a shift in the focusing position of the excitation laser along the longitudinal direction of a nanoantenna (fig.1b) leads to a change in the relative intensity of these maxima. As a result, it is possible to control a switch in the position of a trapped particle between the two distinct tweezing zones with different characteristics (fig.1c-d).

In addition to these features, the peculiar lithography technique used to fabricate the hollow antennas allows us to tune the resonance of the structures on a wide range of different wavelengths without changing the dimension of the inner channel, thus preserving a strong field confinement. Finally, it is worth noticing that it is possible to combine our approach with detection methods both in the visible and near infrared (NIR) range.

**Results and discussion**

*Trapping performances*

Our fabrication technique, based on an established secondary electron lithography technique [19], offers the possibility to tune the wavelength resonance of the nanoantenna independently of the hole dimensions (see fig. 2b) [20]. This approach provides more flexibility with respect to conventional FIB milling of nanoholes into metal substrate, in which the dimension of the plasmonic hot-spot is dictated by resonance conditions. Moreover, the fabrication of sub-10 nm gaps can be hardly achieved with standard techniques.

On the contrary, our method allows us to fabricate few nanometers holes at the tip of the channel, thus strongly increasing the field confinement and opening up the possibility to trap smaller specimens such as DNA or macromolecules.

Numerical simulations in COMSOL Multiphysics have been performed in order to design a hollow nanoantenna that resonates at $\lambda_{exc}$=633 nm.

The choice of a laser wavelength in the low frequency of visible spectrum was due to two main reasons. On one hand, it offers the possibility to excite plasmon resonances in gold antennas, which present stability to oxidation in water, a required feature for tweezing experiments. On the other hand, working with visible wavelength opens the perspective to easily integrate on-line analysis, such as fluorescence or Raman spectroscopy, with a single laser beam configuration. Anyway, it is worth mentioning that the current approach can be shifted to whatever wavelength in both the visible and near-infrared spectral ranges.

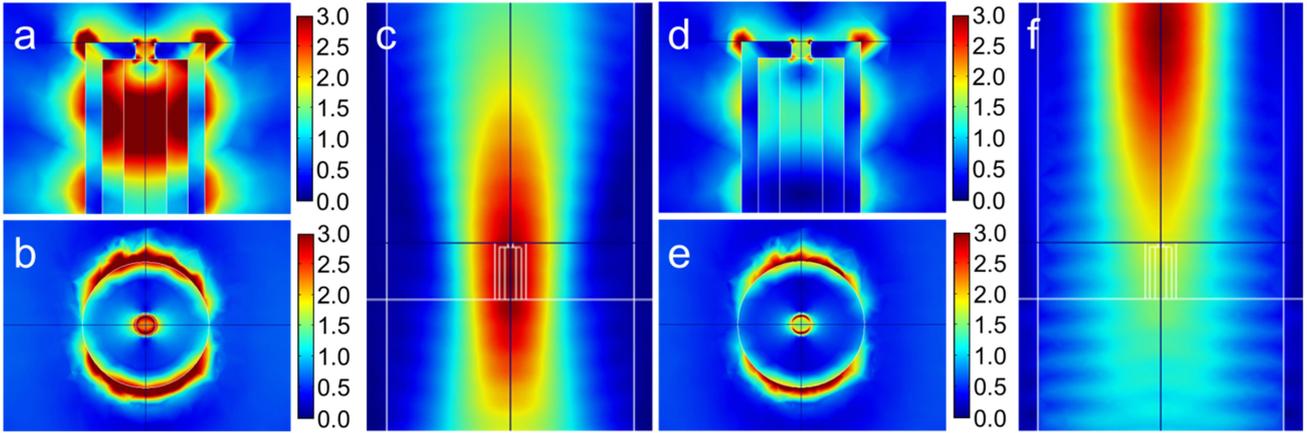

Figure 2. Simulated near-field enhancement distribution of the structure excited at λ = 633 nm with a linearly polarized (before focusing) Gaussian beam with a spot size of 400nm. Side-view (a) and top-view (b) of the |E/E$_0$| field distribution on a nanoantenna when the focus of the exciting beam is located on the antenna tip (c). Side-view (d) and top-view (e) of the EM field distribution when the focus of the exciting beam is located 2 μm above the antenna tip (f).

Figure 2a and 2b respectively show the side-view and top-view of the |E/E$_0$| field distribution for a gold nanoantenna (H=400 nm, D$_e$= 180 nm, D$_i$= 40 nm) irradiated with a Gaussian beam (λ= 633 nm in vacuum) focused on the antenna tip (Fig. 2c). The Gaussian beam is linearly polarized before focusing and it has a spot size of 400 nm at the focal plane. The excitation induces two maxima of field intensity in the inner hole and along the outer wall of the antenna. In this configuration, the field enhancement |E/E$_0$| reaches a maximum value of 2.75 times at the tip of the inner channel, which represents the preferential trapping well.

However, a variation in the irradiation conditions can drastically change the field distribution. Indeed, as reported in Fig. 2(d,e), a shift of 2 μm in the focusing position of the excitation laser along the antenna main axis (Fig. 2f), strongly reduces the spatial extension of the field components inside the channel, while not compromising the distribution of the electromagnetic field on the outer ring of the antenna.

In order achieve a deeper understanding of the two trapping states, we also studied the contribution from the thermal effects generated by the plasmonic heating of the antenna.

Recent literature has revealed that also thermal effects induced by excitation of nanostructures play a substantial role in particle trapping [11,12,13].

Commonly, the illumination of a single antenna generates negligible convection forces [21] and the presence of a heat sink gives additional stability to the tweezing [17].

Nevertheless, accurately predicting the thermal behavior of vertical plasmonic systems is not trivial, since detailed theoretical models have not been developed yet.

Multiphysics simulations of the estimated temperature gradient and of the heat flow generated in both illumination conditions are reported in Figure 3.

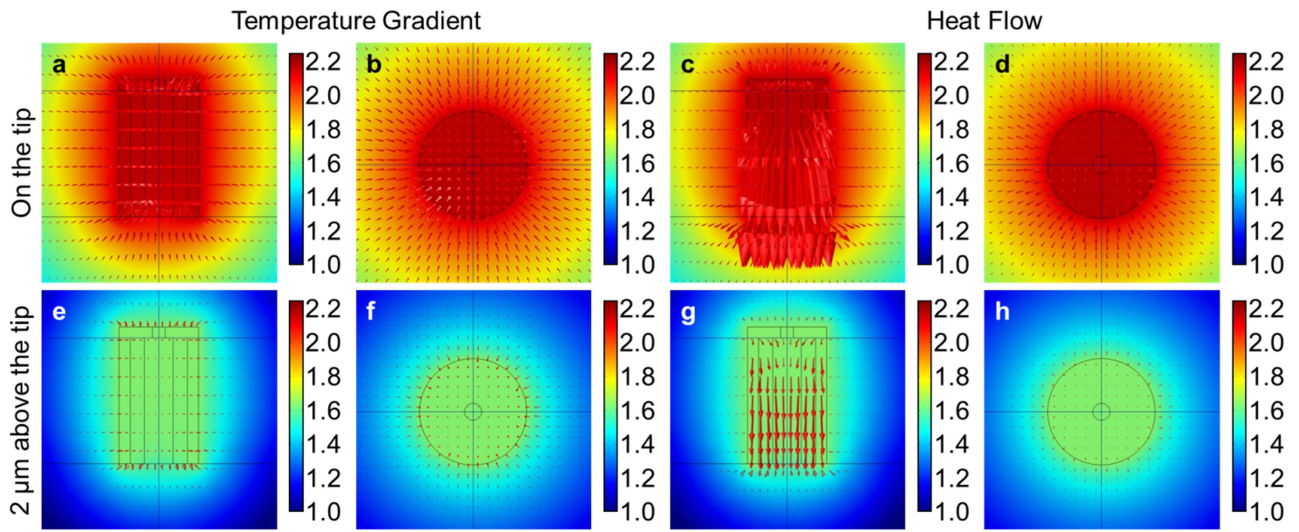

Figure3. On the top row: multiphysics simulations reporting the temperature gradient (a,b) and heat flow (c,d) pointing to the center of the nanochannel for focusing position on the tip on the nanoantenna. On the bottom row, the same parameters reported when the laser is focused 2 μm above the tip of the antenna. While the temperature gradient points to the center of the structure (e,f), the heat flow is directed toward the outer ring of the antenna (g,h).

Fig. 3a,b respectively report the side and top-view of the temperature gradient distribution on the nanochannel when the laser is focused on its tip. The excitation induces a temperature increase of a few degrees, thus confirming the efficient heat dissipation provided by the metal sink at the base of the structure. The gradient presents a uniform intensity along the axis of the antenna and it points towards the center of the channel.

Under these same conditions, the heat flow at the tip of the antenna (reported in fig. 3c,d) also points towards the center of the channel.

By looking at the temperature gradient and at the heat flow, we can state that when the laser is focused on the tip of the antenna, the particles close to the pore are pushed to enter it while the particles at the "outer" ring are pushed out.

When the excitation laser focus is located 2 μm above the structure (fig.3e-h),the excitation of mode inside the pore is less efficient and this has a repercussion on the overall dynamics of the particle.

Indeed, the particle only feels temperature gradients going towards the external part of the pillar and the heat flow pushing it outwards.

Therefore, the particles will be attracted around the edges of the pillar, since the thermal gradient and heat flow are almost negligible at the pore center.

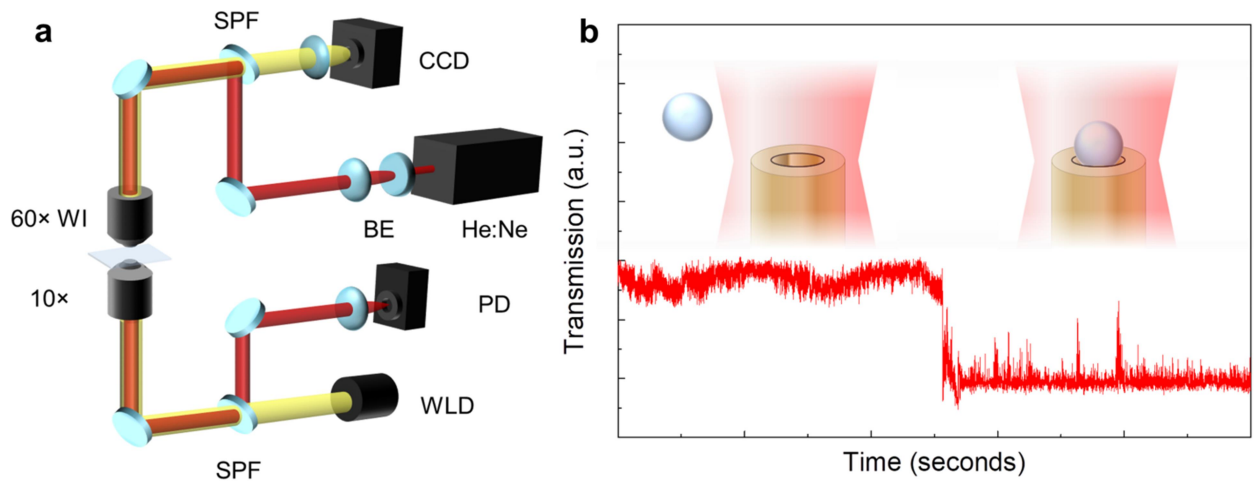

Figure 4. (a) Schematics of the optical setup used for plasmonic tweezing experiments. Legend: 60× WI= 60× Water Immersion Objective N.A. 1.0, 10× O= 10× Objective N.A. 0.3, SPF= Short Pass Filter at 625 nm, BE= Beam Expander, CCD = Detection Camera, He:Ne= 632,8 nm Laser Source, WLD= White Light Illumination Diode, PD= Detection Photodiode.(b) Schematics reporting the change in the transmission when the bead is trapped in the channel

On the base of these results, we performed trapping experiments on the custom-built microscope setup depicted in Fig. 4a. An array of tridimensional antennas was positioned inside a glass-bottom petri dish filled with a suspension of polystyrene nanobeads in water. The array was irradiated through a water immersion objective (NIR WI 60× NA 1.0), which was also used for the imaging by means of a CCD camera. The alignment of the sample was carried out with a 10 nm precision thanks to a piezoelectric controlled XYZ stage.

To induce the plasmonic trapping, a CW fiber-coupled He:Ne laser (λ= 633 nm) was collimated, expanded and focused on the antennas, forming a laser spot with an approximate area of 0.5 µm$^2$.

The laser light transmitted through the sample was collected by a microscope objective (10× NA 0.30) and measured by a silicon-based photodetector. The 10x objective was also used as a condenser to illuminate the sample with cold white light from a LED source, in order to avoid heating effects (see fig. 4a).

The presence of trapping was confirmed by monitoring the change in the transmitted signal through the backside of the membrane (see fig.4b). This change can be attributed to both scattering and fluorescence effects induced by the presence of the dielectric bead.

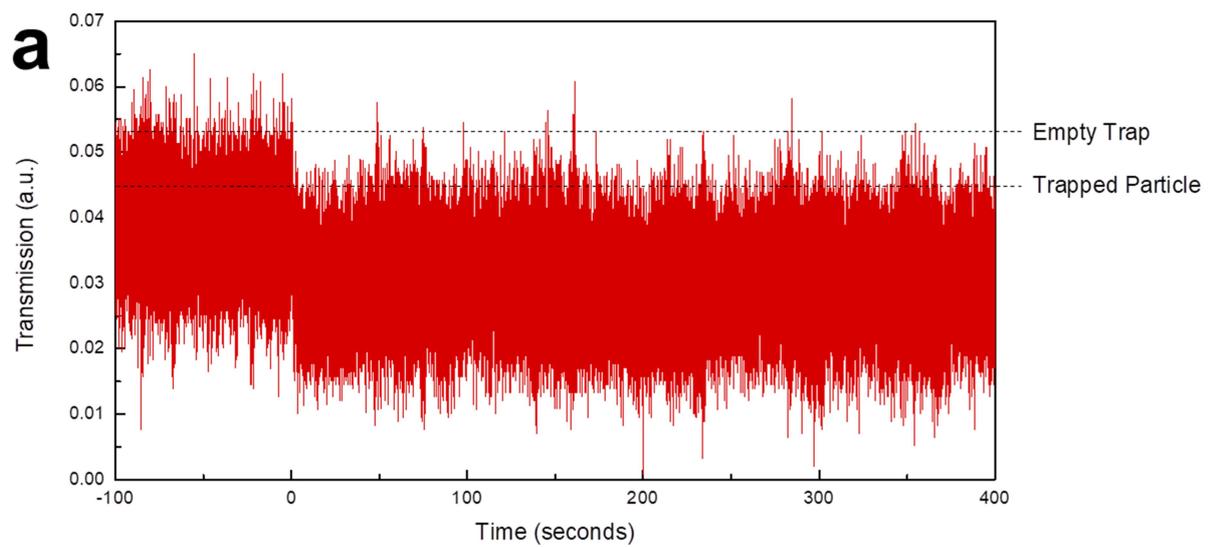

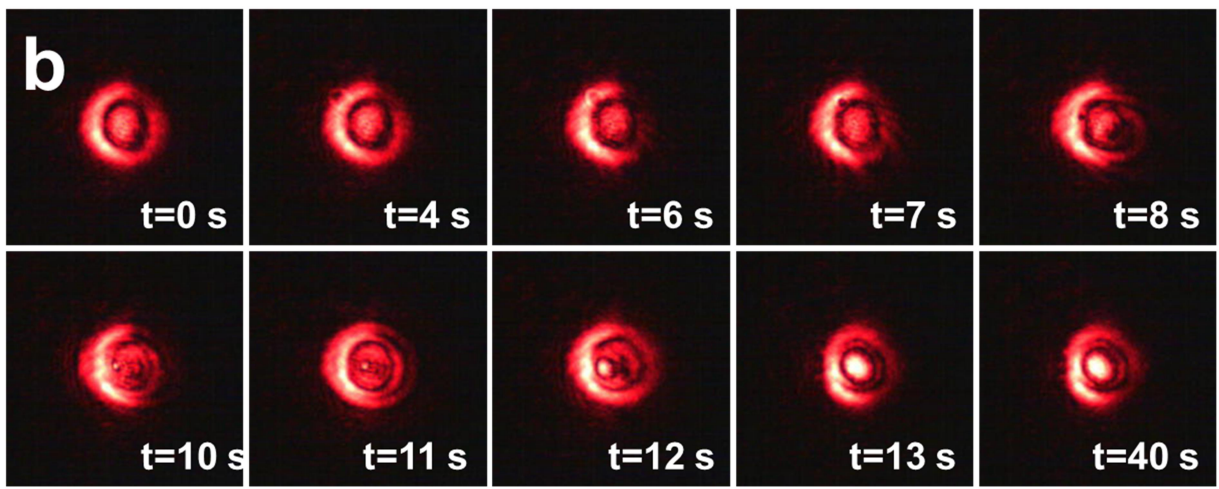

Figure 5. (a) Waveform showing variation of transmitted light as a function of trapping state (b) Temporal sequence showing images on the CCD camera during the trapping event.

Figure 5a reports an example of an experimental time trace obtained irradiating a gold hollow nanoantenna (geometrical parameters: $D_i$= 40 nm, $D_e$= 180, H=400 nm) at λ= 633 nm and demonstrates the optical trapping of a single 100 nm polystyrene bead for a power density of $1 \times 10^9$ W/m$^2$. Under these experimental conditions, we were able to record stable trapping states for up to 5 minutes.

Such data were compared with images collected by the CCD camera and allowed us to relate the highest and lowest transmission levels respectively to an on/off trapping state of a single bead.

In our system, the real time imaging of the trapping event is made possible by the intense scattering of the red laser light off the trapped particle and the antenna's tip. As a result, both the tip of the antenna and the particle are visible on the camera. In this way, we are able to track objects on camera without specific preparation or functionalization, such as the use of fluorescent probes, which require appropriate detection schemes.

In fig. 5b a temporal sequence of CCD images showing the details of a trapping event is displayed. The full video of the trapping event is available in SI (SI_Video 1). In the upper row (t=0-7 s), it is possible to observe the bead approaching the tip of the antenna. Then, the particle enters the potential well and is trapped in the stable position at the aperture of the nanochannel (t=9-40 s).

We would like to underline the fact that 2-3 s are required for the particle to be positioned in the potential minimum (panels t=9-12 s). Such value is in good agreement with the transmission data in Fig. 5a, which report a transition between an on/off trapping state characterized by a few seconds slope.

Measurements were performed on different days with different particle suspensions and showed reproducible results.

No significant variations in the signal intensity were recorded in the absence of nanospheres in the solution, and no optical trapping of the beads was observed when the beam was focused on top of a bead that was out of the reach of the antenna.

## *Switchable trapping*

According to the simulations reported in Fig. 2, a change in the excitation configuration of the nanotube induces a difference in the electric field distribution along the antenna.

In our experiments, this condition has been reproduced by shifting the relative position of the focal point along the main axis of the antenna by 2 µm while trapping a 100 nm particle.

The focal shift resulted in a decreased stability of the trapping position inside the antenna channel, as confirmed by CCD images. This can be observed in SI_video 2, where we show a bead moving away from the top of the antenna, subsequently remaining attached to the outer wall of the nanotube, and then revolving around it

This behavior was reflected in a change of the transmittance levels recorded by the photodiode, as reported in Figure 6.

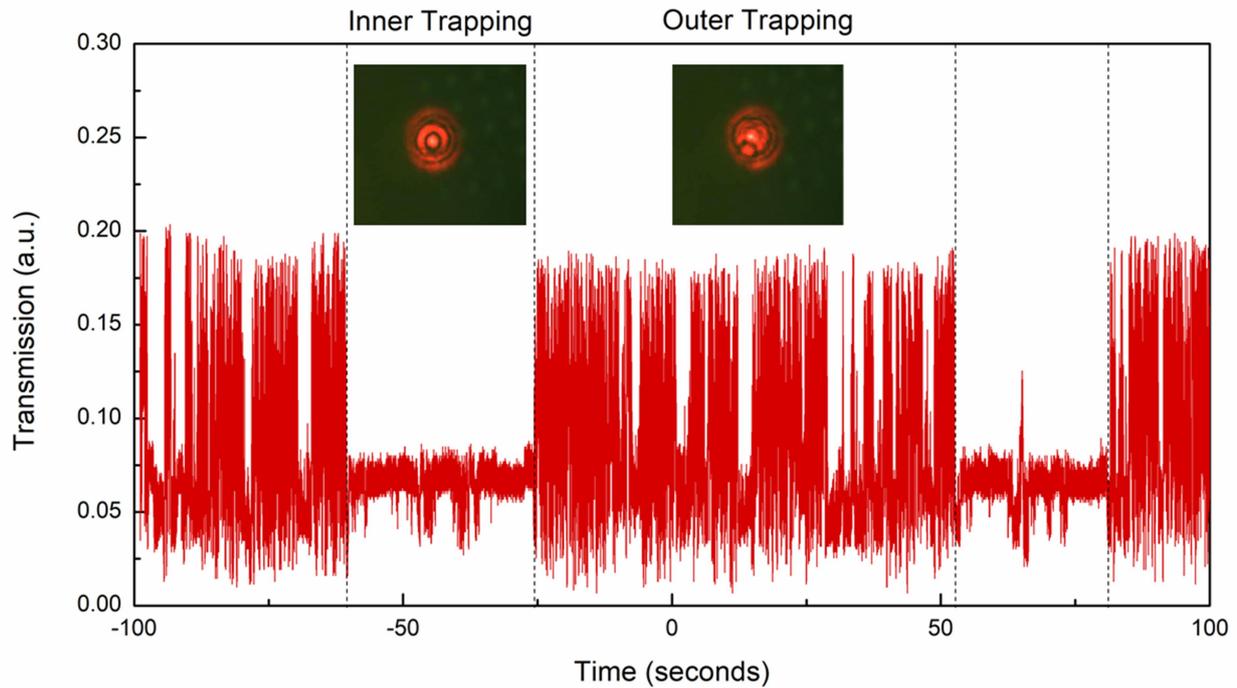

Figure 6. Waveform showing the evidences in the variation of transmitted light as a function of the trapping state. Stronger oscillation in the signal intensity are visible when the particle is trapped on the outer ring of the antenna, due to the dynamic nature of the trapping state.

The transmission signal for the trapping position on the outer wall of the nanotubes ("outer" trapping) showed larger oscillations in the transmission intensity than the trapping state located at the tip ("inner" trapping).

The phenomenon was reproducible and we were able to induce the switch of a 100 nm particle from "inner" to "outer" trapping states and vice versa multiple times (see SI_Video 3).

In the light of the simulations reported in fig.2-3, this effect can be related to a combination of electromagnetic and thermal effects.

Firstly, the electromagnetic simulations show the presence of different field distributions for different excitation configurations.

When the laser is focused on the antenna tip, a combination of high field enhancement and large field extension, which prevents possible particle escaping from the potential well, assure a well localized trapping state.

A shift in the focus of 2 µm above the tip generates a shrink in the field extension, thus limiting the spatial effect of the field enhancement to the inner part of the pore only.

Secondly, the focus change induces an inversion in the heat flow direction around the antenna ring. As a consequence, the particle will be "pushed" towards the center of the antenna for "inner" trapping configuration or towards the outer ring for the "outer" one.

Considering that the nanochannel presents a 40 nm aperture, we can assume that the trapping of the 100 nm particle in the "inner" configuration induces a clogging of the channel aperture.

We would like to remark that the same switch of the focusing position along the antenna did not generate a comparable intensity change in the transmitted light in absence of nanoparticles (see SI Fig.1).

The difference in signal level oscillation can be explained by considering that the intensity in the field enhancement is different in the two described cases. The "outer" trapping position, characterized by a less marked potential depth, will present a less stable condition.

As confirmed by CCD images, indeed, the particle moves around the outer wall of the antenna, in contrast with the "inner" trapping zone which is characterized by a stiffer state.

The difference in stiffness found for the two excitation conditions can be attributed to two main factors. First, as already discussed before, the "outer" position is characterized by a lower field enhancement, thus providing a less stable trapping. Second, it should also be considered that the electric field on the outer wall surface is localized on a way larger area than that of the inner channel, such inducing a revolving movement on the particle.

The combination of these effects lead to a particular trapping state that, in association to recent work in literature [18], can be defined as dynamic. The condition of dynamic tweezing, characterized by a trapped particle moving along a definite pattern, often requires elaborated optical setups. Among these techniques, literature reports works based on polarization modulation [17,22 ], structured beams [18] or double trapping beams [23]

Here, we were able to achieve similar results on a common trapping setup with fixed linear polarization, simply by displacing the microscope objective by 2 µm.

*Trapping efficiency limits*

As a last test, we decided to explore the trapping limits of our platform by performing experiments on smaller particles.

Using the same experimental parameters used for "inner" trapping reported in Fig. 5, we were able to trap 40 nm polystyrene beads on the tip of a plasmonic nanotube.

Nevertheless, whenever the focusing position was switched, the trapping stiffness on the outer wall of the nanoantenna was not strong enough to assure a stable tweezing condition and the particle was released from the trap after some oscillation time (see SI_Video 4).

Such results can be explained by considering the simulations reported in Fig. 2-3.

Indeed, when the antenna is excited on the tip, the EM field presents not only higher values but also higher spatial extension.

Therefore, we can suppose that the most efficient trapping configuration is due to the contributions of high field enhancement and large field extension, which avoids particle escape from the trapping well.

Such combination of effects makes the "inner" trapping state efficient enough to trap particles with dimensions as small as 40 nm.

In principle, it could be possible to further increase the laser power to tweeze smaller particles, thanks to the presence of the heat sink at the base of the nanoantenna that prevents thermal instabilities [24,25].

Nevertheless, we preferred to avoid this approach in view of the potential application of the technique to biological specimens.

**Conclusions**

In this work we demonstrated the possibility offered by tridimensional plasmonic nanotubes to perform the tweezing of 100 nm nano-objects with two different, spatially separated configurations.

The reversible switch between the two conditions can be activated by a variation in the position of the focusing position of the exciting beam along the antenna main axis.

Experimental data have shown that particles can be trapped on the tip of the nanoantenna in a static regime or on the outer wall of the channel in a dynamic regime, presenting a revolving movement.

Multiphysics simulations have revealed that the switch between the two state can be attributed to a change of the EM field distribution and to thermal effects induced by the variation in the laser focus.

The stiffness of the "inner" configuration, the more stable one, is strong enough to allow for the trapping of particles as small as 40 nm.

In principle, this technique is suitable to trap smaller specimens. Indeed, the presence of a heat sink at the bottom of the antenna assures the dissipation of thermal instabilities. Moreover, the fabrication technique offers the possibility to further optimize the structure parameters, such as the pore dimension, in order to achieve higher field enhancements.

The innovative and reproducible two-state switching mechanism opens up novel possibilities in the field of optical manipulation.

For example, particles functionalized with long molecular chains could be attached to the antenna. The switch between trapping positions could offer the opportunity for multiple sequential detections in time.

More in general, the two state trapping systems can represent a model for fundamental studies of plasmonic tweezing mechanisms in dynamic regimes.

Finally, since the trapping laser wavelength is in the visible range, our method can be easily integrate with detection methods such as Raman scattering or fluorescence spectroscopies.

**Acknowledgments**

The research leading to these results was funded by the European Research Council under the European Union's Seventh Framework Programme (FP/2007-2013)/ERC Grant Agreement no. [616213], CoG: Neuro-Plasmonics and the Horizon 2020 Program, Grant Agreement no. [687089], FET-Open: PROSEQO. G.C.M. would like to thank Dr. Andrea Jacassi useful discussion.

**Author Contributions**

G.C.M. conceived the idea of two-state switchable tweezers and fabricated the gold nanochannels. G.C.M. and X.Z-P. built the experimental setup and performed the experiments. N.M. performed the numerical simulations. G.C.M., X.Z-P., D.G., analysed the data. F.D.A supervised the project. All authors contributed to the manuscript preparation.

**References**

[1] Ashkin, A; Dziedzic, J. M. ; Bjorkholm, J. E.;Chu, S; Observation of a single-beam gradient force optical trap for dielectric particles Opt. Lett. **1986,** 11, 288-290

[2] Mestres,P;. Berthelot, J.; Aćimović, S.S.; Quidant, R. Unraveling the optomechanical nature of plasmonic trapping Light – Sci. Appl. **2016**, 5, e16092

[3] Maragò, O.M.; Jones, P.H.; Gucciardi, P.G.; Volpe, G.; Ferrari, A.C.; Optical trapping and manipulation of nanostructures Nat Nanotech.**2013**, 8, 807- 819

[4] Verschueren, D.; Shi, X.; Dekker, C.; Nano-Optical Tweezing of Single Proteins in Plasmonic Nanopores Small Methods. **2019** https://doi.org/10.1002/smtd.201800465

[5] Donato, M.G.; Messina, E.; Foti, A.; Smart, T.J.; Jones, P.H.; Iatì, M.A.; Saija, R.; Gucciardi, P.G.; Maragò, O.M Optical trapping and optical force positioning of two-dimensional materials Nanoscale **2018**, 10, 1245-1255

[6] Grigorenko, A. N.; Roberts, N. W.; Dickinson M. R.; Zhang, Y Nanometric optical tweezers based on nanostructured substrates Nature Photon. **2008**, 2, 365–370.

[7] Righini, M.; Zelenina, A.S.; Girard,C.; Quidant, R.; Parallel and selective trapping in a patterned plasmonic landscape Nat. Phys. **2007**, 3, 477–480

[8] Roxworthy, B. J. ; Ko, K.D.; Kumar, A.; Fung, K.H.; Chow, E.K.C.; Liu, G.L.; Fang, N.X.; Toussaint, K.C. Jr., Application of Plasmonic Bowtie Nanoantenna Arrays for Optical Trapping, Stacking, and Sorting Nano Lett. **2012**, 12, 796–801

[9] Zhang, W.; Huang, L.; Santschi, C.; Martin, O.J.F.; Nano Lett. **2010**, 10, 1006–1011

[10] Sheehan, P. E.; Whitman, L. J. Detection Limits for Nanoscale Biosensors. Nano Lett. **2005**, 5, 803– 807

[11] Ndukaife , J.C.; Xuan, Y.; Nnanna, A. G. A.; Kildishev, A.V.; Shalaev, V.M.; Wereley, S. T.; Boltasseva, A.; High-Resolution Large-Ensemble Nanoparticle Trapping with Multifunctional Thermoplasmonic Nanohole Metasurface ACS Nano **2018**, 12, 5376–5384

[12] Garcia-Guirado, J.; Rica, R. A.; Ortega, J.; Medina, J.; Sanz, V.; Ruiz-Reina, E.; Quidant, R.; Overcoming Diffusion-Limited Biosensing by Electrothermoplasmonics ACS Phot. **2018** *5*, 3673-3679

[13] Kildishev, A.V.; Nnanna, A.G.A.; Shalaev, V.M.; Wereley, S.T.; Boltasseva A. ;. Long-range and rapid transport of individual nano-objects by a hybrid electrothermoplasmonic nanotweezer. Nat. Nanotech. **2016**,11,53–59

[14] Juan, M.L.; Gordon, R.; Pang, Y.; Eftekhari F.; Quidant, R. Self-induced back-action optical trapping of dielectric nanoparticles Nat. Phys. **2009**, 5, 915 - 919

[15] Pang, Y.; Gordon, R.; Optical Trapping of 12 nm Dielectric Spheres Using Double-Nanoholes in a Gold Film Nano Lett. **2011**, 11, 3763–3767

[16] Berthelot, J.; Aćimović, S. S.; Juan, M. L.; Kreuzer, M. P.; Renger, J.; Quidant, R.; Three-dimensional manipulation with scanning near-field optical nanotweezers Nat. Nanotech. **2014** 9, 295–299

[17] Wang, K.; Schonbrun, E.; Steinvurzel, P.; Crozier, K.B., Trapping and rotating nanoparticles using a plasmonic nano-tweezer with an integrated heat sink, Nat. Comm. **2011**, 2, 469

[18] Huft, P. R.; Kolbow, J.D.; Thweatt, J.T.; Lindquist, N.C.; Holographic Plasmonic Nanotweezers for Dynamic Trapping and Manipulation Nano Lett. **2017**, 17, 7920–7925

[19] De Angelis, F.; Malerba, M.; Patrini, M.; Miele, E.; Das, G.; Toma, A.; Zaccaria, R. P.; Di Fabrizio, E. 3D Hollow Nanostructures as Building Blocks for Multifunctional PlasmonicsNano Lett. **2013**, 13, 3553– 3558,

[20] Malerba, M.; Alabastri, A.; Miele, E.; Zilio, P.; Patrini, M.; Bajoni, D.; Messina, G.C.; Dipalo, M.; Toma, A.; Proietti Zaccaria, R.; De Angelis, F;. 3D vertical nanostructures for enhanced infrared plasmonics Scientific Reports **2015**, 5, 16436

[21] Donner, J. S.,;Baffou, G.; McCloskey, D.; Quidant, R. Plasmon-Assisted Optofluidics ACS Nano **2011**, 5, 5457–5462.

[22] Tsai, W.Y.; Huang, J.-R.; Huang, C-B.; Selective Trapping or Rotation of Isotropic Dielectric Microparticles by Optical Near Field in a Plasmonic Archimedes Spiral Nano Lett. **2014**, 14, 547–552

[23] Zhang, Y.; Shen, J.; Xie, Z.; Dou, X.; Min, C.; Lei, T.; Liu, J.; Zhuc, S.; Yuan, X.; Dynamic plasmonic nano-traps for single molecule surface-enhanced Raman scattering Nanoscale **2017**,9, 10694-10700

[24] Messina, G.C.; Dipalo, M.; La Rocca, R.; Zilio, P.; Caprettini, V.; Proietti Zaccaria, R.; Toma, A.; Tantussi, F.; Berdondini, L.; De Angelis F.; Spatially, Temporally, and Quantitatively Controlled Delivery of Broad Range of Molecules into Selected Cells through Plasmonic Nanotubes Adv.Mater.**2015**, 27, 7145–7149

[25] Mancini, A.; Giliberti, V.; Alabastri, A.; Calandrini, E.; De Angelis, F.; Garoli, D.; Ortolani, M. Thermoplasmonic Effect of Surface-Enhanced Infrared Absorption in Vertical Nanoantenna Arrays J. Phys. Chem. C **2018**, 122, 13072–13081

# Switchable two-state plasmonic tweezers for dynamic manipulation of nano-objects

Gabriele C. Messina[1,2*], Xavier Zambrana-Puyalto[1], Nicolò Maccaferri[1], Denis Garoli[1], Francesco De Angelis[1*]

[1]Plasmon Nanostructures, Istituto Italiano di Tecnologia, Via Morego 30, 16163 Genova GE, Italy

[2]Present Address: Intense Laser Irradiation Laboratory (ILIL), Istituto Nazionale di Ottica (CNR-INO), Via Giuseppe Moruzzi 1, 56124 Pisa PI, Italy

*Corresponding author: gabriele.messina@ino.it; francesco.deangelis@iit.it

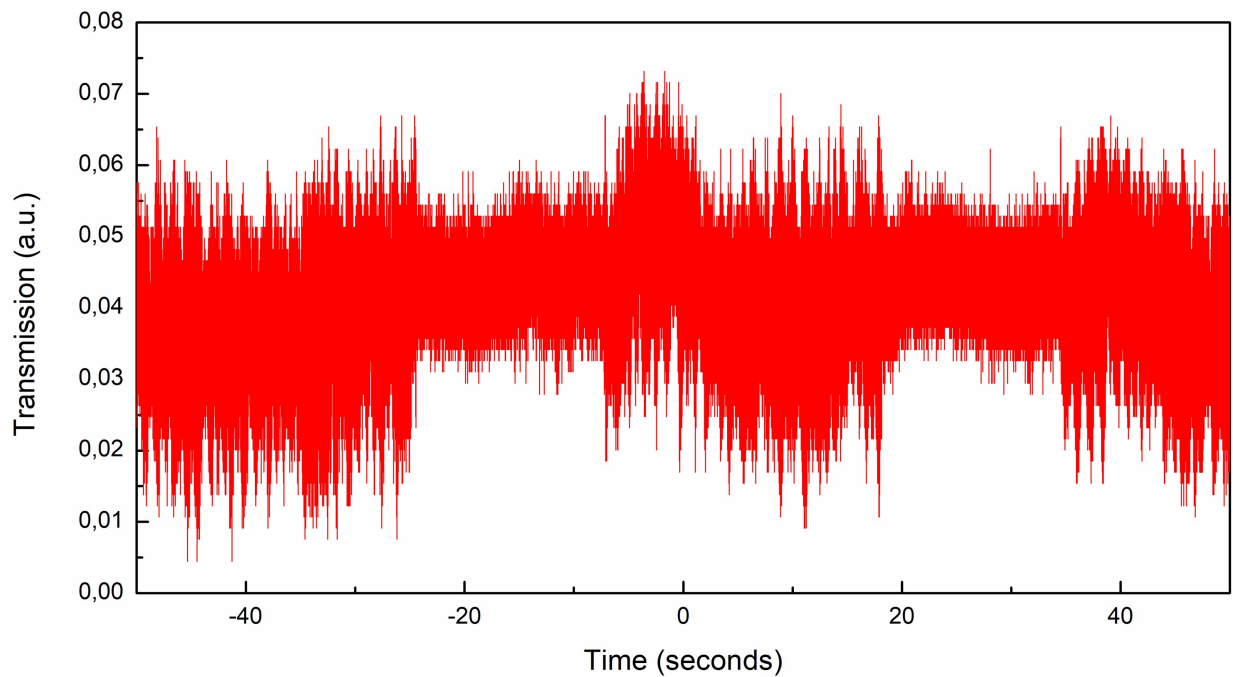

Figure 1. Change in the transmittance as function of the focusing position in absence of trapped bead.